\documentclass[]{article}

\usepackage{amsmath, amsthm, amssymb, amsfonts}
\usepackage{mathtools}
\usepackage{hyperref}
\usepackage{graphicx}
\usepackage{dsfont}
\usepackage{fullpage}
\usepackage{color}
\usepackage{soul} 
\usepackage[title]{appendix}
\usepackage{tikz}
\usetikzlibrary{patterns}
\usetikzlibrary{shapes.multipart}
\usetikzlibrary{arrows}

\theoremstyle{definition}
\newtheorem{remark}{Remark}[section]

\definecolor{labelkey}{cmyk}{.4,.2,0,0}

\newcommand{\be}{\begin{equation}}
\newcommand{\ee}{\end{equation}}
\newcommand{\bea}{\begin{eqnarray}}
\newcommand{\eea}{\end{eqnarray}}

\newcommand{\R}{{\mathbb R}}

\newcommand{\ssp}{\hspace{3pt}}

\newcommand{\I}{\mathbf i}

\usepackage{titlesec}
\titleformat{\section}{\large\bf}{\thesection}{1em}{}
\titleformat{\subsection}[runin]{\bf}{\thesubsection}{1em}{}[.]

\usepackage{authblk}
\title{This is some thing}
\author[1]{Guillaume Barraquand and Pierre Le Doussal}
\affil[1]{\normalsize Laboratoire de Physique de l'\'Ecole Normale Sup\'erieure, ENS, Universit\'e PSL, CNRS, Sorbonne Universit\'e, Universit\'e de Paris, 75005 Paris, France}

\title{\bf \large Moderate deviations for diffusion in time dependent random media}
\date{}

\begin{document}

\maketitle

\begin{abstract}
	The position $x(t)$ of a particle diffusing in a one-dimensional uncorrelated and time dependent random medium is simply Gaussian distributed in the typical direction, i.e. along the ray $x=v_0 t$, where $v_0$ is the average drift. However, it has been found that it exhibits  at large time sample to sample fluctuations characteristic of the KPZ universality class when observed in an atypical
	direction, i.e. along the ray $x = v t$ with $v \neq v_0$. Here we show, from exact solutions, that in the moderate deviation regime $x - v_0 t \propto t^{3/4}$ these fluctuations
	are precisely described by the finite time KPZ equation, which thus describes the crossover 
	between the Gaussian typical regime and the KPZ fixed point regime for the large deviations. 
	This confirms 
	heuristic arguments given in   \cite{PLDTTDiffusion}. These exact results include the discrete model known as the Beta RWRE, and a continuum diffusion. They predict the behavior of the maximum of a large number of independent walkers, which should be easier to observe (e.g. in experiments) in this moderate deviations regime.
\end{abstract}


\section{Introduction}

It has been found that diffusion in uncorrelated time dependent random environments, observed away from
the typical direction, exhibits  at large time the behavior of the Kardar-Parisi-Zhang (KPZ) universality class.  Specifically, and restricting here to one space dimension $d=1$, the typical behavior, i.e. in the direction of the
drift $v_0$ along the  space time ray $x = v_0 t$, is simple Gaussian diffusion.
However, if one looks along the ray $x = v t$, with $v \neq v_0$, the sample to sample fluctuations (i.e. fluctuations due to the randomness of the  environment) 
of the logarithm of the probability distribution are described by the Tracy Widom distribution, characteristic of the KPZ class. This has been first discovered, and shown rigorously, for a discrete model of random walk in a time dependent random environment (RWRE) known as the Beta RWRE \cite{barraquand2017random}. 
 This result concerns the far tail of the probability distribution, in the so-called large deviation regime.
  In other terms,  if one considers a sufficiently large collection of $N$ independent random walkers in the same environment, with $\log N \propto t$,  the Tracy-Widom distribution also describes the sample to sample fluctuations of the rightmost walker. In principle, although quite challenging, this may be detectable in experiments on e.g. tracer diffusion in fluids. 
\bigskip 

\begin{figure}
	\begin{center}
		\begin{tikzpicture}[scale=1.5]
		\draw[thick, -> ] (-4,0) -- (5,0);
		\draw[thick, -> ] (0,0) -- (0,1.8) node[anchor = north east] {PDF of $x(t)$};
		\draw[dashed, gray] (1,-1) -- (1,1.2);
		\draw[dashed, gray] (3,-1) -- (3,1.2);
		\draw[dashed, gray] (-1,-1) -- (-1,1.2);
		\draw[dashed, gray] (-3,-1) -- (-3,1.2);
		\draw[thick,  domain=-4:4.9, samples=200] plot(\x, {exp(-((\x)^2))+0.1+0.02*sin(3000*\x)-(\x)^2/350});
		\begin{scope}[yshift=-0.3cm]
		\draw  (0,0) node[text width=1.8cm,align=center] {$x\propto t^{1/2}$};
		\draw  (2,0) node[text width=1.8cm,align=center] {$x\simeq \tilde x t^{3/4}$};
		\draw  (4,0) node[text width=2cm,align=center] {$x\simeq \tilde x t$};
		\end{scope}
		\begin{scope}[yshift=-1.1cm]
		\node[draw,rectangle, rounded corners=3pt,fill=gray!10,text width=2.2cm,align=center] () at (0,0) {Gaussian typical regime};
		\node[draw,rectangle, rounded corners=3pt,fill=gray!10,text width=1.8cm,align=center] () at  (2,0) {Moderate deviations};
		\node[draw,rectangle, rounded corners=3pt,fill=gray!10,text width=1.7cm,align=center] () at  (4,0) {Large deviations};
		\end{scope}
		
		\draw (0.5,1.3) node[] {$e^{\frac{-x^2}{4t}+ \rm{EW}}$};
		\draw (2,0.7) node[] {$e^{-\frac{ \tilde x^2}{4t} t^{1/2}+ \rm{KPZ}}$};
		\draw (4,0.5) node[] {$e^{- J(\tilde x) t + \rm{TW}}$};
		\end{tikzpicture}
	\end{center}
	\caption{Schematic plot of the PDF of the distribution of the diffusion $x(t)$ (we assume that the average drift $v_0=0$ for simplicity). At first sight this is a Gaussian curve, with some roughness because of sample to sample fluctuations. There are three distinct regimes, symmetric with respect to the origin. For each one, we indicate the order of decay of the PDF as well as the nature of the sample to sample subleading fluctuations: Edwards-Wilkinson (EW) in the Gaussian typical regime, Tracy-Widom distributed (TW) in the large deviations regime, and distributed as the KPZ equation (KPZ) in the moderate deviations regime, which is the main focus of this paper.}
	\label{fig:plotPDF}
\end{figure}
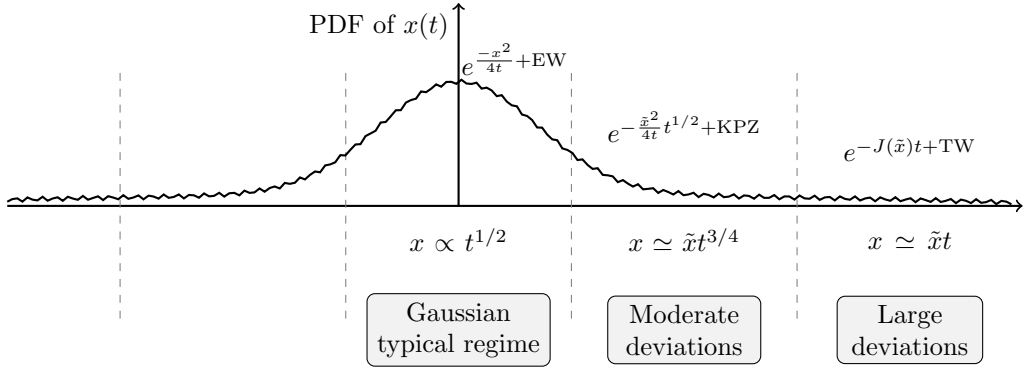

\noindent A quite simple physical argument was given in \cite{PLDTTDiffusion} for a broad connection (in any space dimension $d$) between continuum diffusions in random environments and the KPZ equation. Let us recall that the 
KPZ equation \cite{KPZ} (here restricted in $d=1$) was introduced to describe the stochastic growth of an interface parameterized by a height field $h(x,t)$, $x \in \mathbb{R}$, as a function of time $t$ and reads
\be \label{kpz1} 
\partial_t h = \nu_0 \partial_x^2 h + \frac{\lambda_0}{2} (\partial_x h)^2 + \sqrt{D_0} \eta(x,t),
\ee
where $\nu_0$ is diffusivity, $\lambda_0$ the non-linearity and $D_0$ measures the amplitude of
the noise. Here $\eta$ is the noise and the case of main interest is 
when $\eta$ is a unit space-time white noise. The KPZ equation \eqref{kpz1} describes the crossover between the simpler linear Edwards-Wilkinson equation 
(i.e. \eqref{kpz1} with $\lambda_0=0$) at short time $t \ll t^*$
and the KPZ fixed point at large time $t \gg t^*$. When $\eta$ is unit space-time white noise
the crossover time scale 
is given by $t^*=  2 (2 \nu_0)^5/(D_0^2 \lambda_0^4)$ and one defines a corresponding
length scale $x^*=(2 \nu_0)^3/(D_0 \lambda_0^2)$. 
The KPZ fixed point describes the universal large time behavior of all models in the KPZ class. 
It was argued in   \cite{PLDTTDiffusion}  that not only the KPZ fixed point, but the KPZ equation itself, hence the EW to KPZ crossover, should be observable in the regime of moderate deviations of the diffusion (see Figure \ref{fig:plotPDF}).
This regime corresponds to $x - v_0 t \propto t^{3/4}$, i.e. in the tail, but closer to the typical 
direction than the aforementioned large deviation regime. 
The argument, recalled in details in Section \ref{sec:physics}, uses a transformation of the diffusion problem
into an equation identical to the KPZ equation up to some additional terms, and assuming
that these terms are RG irrelevant above a certain (small) length and time scale. Since it was 
somewhat heuristic, it seems useful to obtain confirmation and exact results about the
moderate deviation regime. This, and exploring some consequences, is the aim of the present paper. One of the consequences is a precise estimate of the position of the maximum of
$N$ walkers, when $\log N$ is now scaled as $t^{1/2}$. Since it requires a smaller number of
walkers  than in the large deviation regime, this suggests that  the moderate deviation regime will be easier to detect in experiments.

\bigskip 
\noindent Besides the results of \cite{barraquand2017random, PLDTTDiffusion} let us mention a few other relevant works. First, the KPZ equation also arises from random walks or diffusions under a different scaling: in the large deviation regime, but when the noise amplitude parameter is scaled to $0$, the KPZ equation appears.  This was proved in \cite{corwin2017kardar} for a large class of discrete random walks in space-time random environments and in \cite{barraquand2019large} for an integrable model of diffusions in random environment \footnote{The result was proved in \cite{barraquand2019large} in terms of convergence of moments, while it was proved in a much stronger sense in \cite{corwin2017kardar}, i.e. weak convergence in the space of continuous functions of space and time. In the present paper, we will ignore these mathematical aspects.}. The fact that the KPZ equation arises from different scalings should not be a surprise. It is well-known that models in the KPZ class possessing tunable parameters controlling the strength of the noise or the amplitude of the non linearity should converge to the KPZ equation when the model is rescaled and these parameters are appropriately tuned. This phenomenon is usually referred to as weak universality. The result of \cite{corwin2017kardar} corresponds to the so-called  \emph{weak noise scaling} of random walks in random environment, while we investigate in this paper the \emph{weak non linearity scaling} (sometimes called weak asymmetry scaling).  
In the typical direction, i.e. along the ray $x = v_0 t$, we have mentioned that the behaviour is Gaussian. This was proved rigorously in the mathematics litterature, see e.g. \cite{rassoul2009almost} and references therein. The subleading sample to sample fluctuations are  described by the Edwards-Wilkinson universality class (i.e. \ref{kpz1} with $\lambda_0=0$). This was proved in \cite{RAP} for discrete models of random walks in random environment in dimension $1+1$, and in \cite{Yu16} for a certain class of continuous diffusions which we discuss below in Section \ref{sec:stickyBM}. In the large deviation regime (i.e along the ray $x=vt$ with $v\neq v_0$), the fact that the probability distribution admits a large deviation principle for almost every environment was proved in \cite{LargeDev} in a quite general setting. It was then shown in \cite{barraquand2017random}, using the exactly solvable Beta RWRE, that the second order corrections to the large deviation principle fluctuate sample to sample according to the Tracy-Widom distribution, as we have already mentioned. Note that the exact solvability of the Beta RWRE is rooted in a work of Povolotsky \cite{povolotsky2013integrability} on Bethe ansatz solvable probabilistic models of interacting particles.  The connection between the Beta RWRE and the KPZ universality class was further strengthened and refined in \cite{usBeta, balazs2019large}.

\subsection*{Outline of the paper} In this paper, we first recall in Section \ref{sec:physics} the physical arguments  from \cite{PLDTTDiffusion}. We then consider in Section \ref{sec:integrablediffusions} an exactly solvable model of one-dimensional diffusions in random environment. This model was first introduced by Le Jan and Raimond  \cite{le2004flows} using the language of stochastic flows (see also \cite{le2004products, le2004sticky}) and further studied in \cite{Howitt2009consistent, schertzer2014stochastic} in connection to families of so-called  sticky Brownian motions and the Brownian web/net (see the review \cite{schertzer2015brownian}). The exact distribution of the probability distribution of a diffusion in this model (conditionally on the random environment) was computed exactly in \cite{barraquand2019large}. We use the exact formulas from \cite{barraquand2019large} to prove that the fluctuations of the probability distribution in the regime  $x - v_0 t \propto t^{3/4}$ are described by the KPZ equation, thereby confirming the heuristic arguments from Section \ref{sec:physics}. More precisely, we show that the moments of the cumulative distribution function (CDF) of the random diffusion (with possibly varying starting points), when appropriately scaled, converge to the (exponential) moments of the KPZ equation. In Section \ref{sec:RWRE}, we prove an analogous statement on another exactly solvable model, the Beta random walk, introduced in \cite{barraquand2017random}. In Section \ref{sec:maximum}, we state the main consequences of our results concerning the maximum of $N= e^{c\sqrt{t}}$ diffusions $\lbrace x_i(t)\rbrace_{i=1, \dots, N}$ drawn independently in the same environment: the position of the maximum of these diffusions has fluctuations given by the sum of a Gumbel random variable  and a random variable distributed as the solution to the KPZ equation. Finally, in  \ref{sec:interpolation}, we consider a continuous model interpolating between the KPZ equation and the model from Section \ref{sec:integrablediffusions}. It is worth noticing that this interpolation is still Bethe ansatz solvable in a similar way as in \cite{barraquand2019large} (this fact was hinted in earlier works, in particular \cite{stouten2018something}), but this model does not introduce new physics since it reduces to the model from Section \ref{sec:integrablediffusions} after appropriate scalings and changes of variables. 

\subsection*{Acknowledgements} G.B. thanks Mark Rychnovsky for useful conversations regarding sticky Brownian motions  and for drawing his attention to the reference \cite{warren2015sticky}. PLD is grateful to Thimoth\'ee  Thiery for previous collaborations related to this work. We are grateful to Denis Bernard for helpful remarks and pointing out some references.  We also thank Eric and Ivan Corwin for discussions. PLD acknowledges support from ANR grant ANR-17-CE30-0027-01 RaMaTraF.

\section{First model: physical argument and length scales}
\label{sec:physics} 

In this Section we recall and summarize the physical argument given in   \cite{PLDTTDiffusion}
and the discussion of the important length scales. 
It was given in arbitrary space dimension $d$, but here we will discuss it only in $d=1$. Consider the following Langevin equation which describes the position $x(t)$ of a particle (also called a walker below)
in the time dependent random force field $\xi(x,t) + v_0$ (called the environment), with
$\mathbb E\left[{\xi(x,t)}\right]=0$ and a bias $v_0$ 
\bea \label{langevin} 
\frac{d}{d t} x(t) = \xi(x(t),t) + v_0 + \sqrt{2 D_b} \, \zeta(t) ,
\eea 
with $\zeta$ a (thermal) unit Gaussian white noise, $\langle \zeta(t)   \zeta(t') \rangle= \delta(t-t')$, and $D_b$ the bare diffusion coefficient. Here and below $\langle \cdot \rangle$ refers to the average over thermal fluctuations $\zeta$, $\mathbb E[\cdot]$ the average over the environment $\xi(x,t)$, and ``sample to sample'' fluctuations refer to fluctuations over the environment. For definiteness we will consider $\xi(x,t)$ a smooth Gaussian random field with short-range correlations
in space (over scale $r_c$) and time (over scale $\tau_c$), 
\bea \label{Defr0}
\mathbb E\left[{\xi(x,t)  \xi(x',t')}\right] = \sigma  \, \delta_{r_c}(x-x') \delta_{\tau_c}(t-t') \ssp  , \quad  \quad \text{Model I},
\eea  
which together with \eqref{langevin} defines model I. 
Here $r_0$ has dimension of a length. Here $\delta_a(x)$ denotes a mollified delta function,
i.e. $\delta_a(x) = a^{-1} g(x/a)$ for some choice of a rapidly decaying function $g(x)$
with $\int dx g(x)=1$. We also define a version of the model, which we call model II,
such that the correlation time $\tau_c$ is taken to be zero, i.e for an environment delta correlated in time, with Ito convention
\be \label{Model2} 
\mathbb E\left[{\xi(x,t)  \xi(x',t')}\right] = \sigma  \, \delta_{r_c}(x-x') \delta(t-t') \ssp   ,\quad  \quad \text{Model II}.
\ee 
The associated Fokker-Planck equation for the probability
distribution function (PDF), $P(x,t) = \langle \delta(x(t)-x) \rangle$, of the position $x$ of the particle at time $t$ in a given environment $\xi$ reads
\be \label{bk} 
\partial_t P = D \partial_x^2 P - \partial_x [ (\xi(x,t) + v_0) P ].
\ee
In the model I where $\xi(x,t)$ is smooth in time, \eqref{bk} holds with $D=D_b$, however 
in the model II, the diffusion coefficient is dressed by the small time scale
fluctuations of the random field, and from the Ito rule one obtains $D= D_b + \sigma \delta_{r_c}(0)$. Note that $\sigma/D$ defines a length scale, called $r_0$ in \cite{PLDTTDiffusion}
(this renormalisation effect of $D$ was not taken into account there).

One then performs a change of variable which is motivated as follows. In model II, i.e. with $\tau_c=0$, the sample averaged PDF is simply the Gaussian $\mathbb E\left[{P(x,t)}\right] = \frac{1}{\sqrt{4 \pi t}} e^{- \frac{(x- v_0 t)^2}{4 D t}}$ with a drift velocity 
$v_0$. This actually also describes the typical walk (i.e. in the direction of the drift) in the typical environment. We are interested in looking in the direction $x=v t + o(t)$, away from the typical direction $x=v_0 t$. We denote $u=v-v_0$ the difference between the observation direction and the typical direction. It means that we are probing the tail of the PDF, and
it is thus natural to factor out the main dependence in $x$ (which is exponential) in that region, and perform the change of variable
\be
P((v_0+u) t + x,t) = e^{- \frac{x u}{2 D} - \frac{u^2}{4 D} t } \hat Z(x,t) .
\ee
Then $\hat Z(x,t)$ satisfies 
\bea \label{she1} 
\partial_t \hat Z(x , t) =  D \partial_x^2 \hat Z(x , t)   +  \frac{u}{2 D} \hat \xi(x,t)
\hat Z - \mu \, \partial_x (\hat \xi \hat Z)
\eea  
with $\mu=1$. Here $\hat \xi(x,t)=\xi(x+(v_0+u) t,t)$ is a Gaussian noise
with the same correlations as $\xi(x,t)$ (for Model II).
The equation with $\mu=0$ is the stochastic heat equation (SHE), which
is related to the KPZ equation \eqref{kpz1} with $\nu_0=\frac{\lambda_0}{2}=D$,
via the Cole-Hopf transform $h(x,t)= \log \hat Z(x,t)$. 
In \cite{PLDTTDiffusion} (see also Supp. Mat. there) it was argued that the additional term $\partial_x (\xi \hat Z)$ in \eqref{she1}, since it contains additional derivatives, is irrelevant by power counting above a certain scale. This scale was estimated to be the diffusion scale $x= x_0 = D/u$ and $t=t_0=x_0^2/D=D/u^2$
(above this scale the free diffusion starts to feel the bias $u$). Hence above this scale, and in the region $x = o(t)$, $\hat Z(x,t)$ should behave as the solution of the SHE
with a noise amplitude $\propto u/(2D)$, and $h(x,t)$ as the corresponding solution of the KPZ equation. This implies KPZ fixed point behavior at large time for
the diffusion. It predicts e.g. a GUE-TW distribution for the sample to sample fluctuations of $\log P(x,t)$
when properly centered and scaled by $t^{1/3}$, for an initial condition (IC) $P(x,t=0)$ localized
in space. It predicts a GOE-TW distribution for the IC $P(x,0) \propto e^{- \frac{x u}{2 D}}$. Such a behavior was  confirmed numerically for a lattice version of the model in \cite{PLDTTDiffusion}.\\

The above arguments predict more, i.e. they predict a crossover from EW to KPZ behaviors. For small enough $r_c$, one can estimate the scales
of the  EW-KPZ crossover by using those of the KPZ equation with white noise. For the 
model II the parameters of the associated
KPZ equation are $D_0=u^2 r_0/(4 D)$, $\nu_0=\frac{\lambda_0}{2}=D$. 
The scales of the EW-KPZ crossover are then estimated as $x^*=(2 \nu_0)^3/(D_0 \lambda_0^2) = \frac{8 D^2}{r_0 u^2}$,
$t^* = 2 (2 \nu_0)^5/(D_0^2 \lambda_0^4)=(4 D)^3/(r_0^2 u^4)$. 
Hence if $t^* \gg t_0$, that is for {\it small enough difference} $u=v-v_0$ with the typical direction, more precisely for $u \ll 8 D/r_0$, 
the PDF of $\log P$ should be described by the finite time KPZ equation. The crossover from
EW to KPZ should thus be observable, above the scales $x_0$ and $t_0$
(see Supp. Mat.  of \cite{PLDTTDiffusion} p. 13-14). 
It was thus predicted that, for small enough $u$ and for $x \gg x_0$, $t \gg t_0$ it is reasonable to expect that
\be \label{approx} 
\log P( (v_0 + u) t + x, t) \simeq 
- \frac{u^2 t}{4 D} - \frac{x u}{2 D} + h^{\rm KPZ}\left(\frac{x}{x^*},\frac{t}{t^*}\right)  
 ,\quad \quad x^*= \frac{8 D^2}{r_0 u^2}
 ,\quad \quad t^* =\frac{(4 D)^3}{r_0^2 u^4},
\ee 
where here and below we denote $h^{\rm KPZ}(x,t)=h(x,t)$ the solution of the KPZ equation
in the reduced units defined by
\be \label{KPZh} 
\partial_t h = \partial_x^2 h + (\partial_x h)^2 + \sqrt{2} \, \eta(x,t),
\ee 
where $\eta(x,t)$ is a unit space-time white noise (in these units $x^*=t^*=1$). 
The proper definition of the solution is via the Cole-Hopf formula $h(x,t)= \log Z(x,t)$ 
where $Z(x,t)$ satisfies the SHE
\be \label{she} 
\partial_t Z = \partial_x^2 Z + \sqrt{2} \eta(x,t) Z .
\ee 
The prediction \eqref{approx} is expected to be valid as a process, i.e. as $x$ is varied within $x=o(t)$,
and for a large set of initial conditions (which should include the droplet, flat and Brownian initial conditions considered usually). We discuss just below and in Section \ref{sec:LeJanRaimondflow} several initial conditions $P(x,0)$ which lead to droplet initial data for $h^{KPZ}$.

The argument in \cite{PLDTTDiffusion} goes on by considering the droplet IC for the KPZ equation, which corresponds to
an initial particle at $x(0)=0$ for the diffusion.
The prediction \eqref{approx} for the one-point PDF (setting $x=0$) 
interpolates between two limits, KPZ fixed point and Gaussian:

(i) For fixed (and small) $u$ and large $t \gg t^* \propto 1/u^4$
we can use that $h^{\rm KPZ}(0,t/t^*) = - \frac{t}{12 t^*} + (\frac{t}{t^*})^{1/3} \chi_2$ 
where $\chi_2$ is a GUE-TW random variable, and obtain
\be \label{res2}
\log P((v_0 + u)t,t)  \simeq - J(u) t + \lambda(u) t^{1/3} \chi_2  , \quad \quad J(u) \simeq \frac{u^2}{4 D}  + \frac{2 r_0^2 u^4}{3 (8 D)^3}  ,\quad \quad \lambda(u) \simeq \frac{r_0^{2/3} u^{4/3}}{4D}, 
\ee
an estimate which should be valid in the weak bias limit $u = O(1) \ll D/r_0$ of the large deviation regime
(for which the KPZ fixed point behavior extends to any $u \neq 0$). This is very reminiscent of
the results obtained in \cite{barraquand2017random} for the Beta polymer (see discussion below). 

(ii) If we set $u={\sf x}/t$, and consider $t \ll t^*$ small, the first term 
in \eqref{approx} dominates, and one recovers the Gaussian 
$P( v_0 t + {\sf x}, t) \simeq e^{- \frac{{\sf x}^2}{4 D t}}$.\\ 

In between these two limits, an intermediate regime was identified, by
comparing $t$ and $t^* \propto 1/u^4 = t^4/{\sf x}^4$, i.e. ${\sf x} \propto t^{3/4}$. For ${\sf x} = u t = x - v_0 t = O(t^{3/4})$,
it was argued from \eqref{approx} that
\be \label{predict} 
\log P(v_0 t + {\sf x},t) \simeq - \frac{{\sf x}^2}{4 D t} + h^{\rm KPZ}(0, \tilde {\sf x}^4)  , \quad \quad \tilde {\sf x} = \frac{{\sf x}}{r_0 (\frac{4 D t}{r_0^2})^{3/4}},
\ee 
which describes the EW to KPZ crossover. In the EW regime $\tilde {\sf x} \ll 1$ the second term
is distributed as $c_0 \tilde {\sf x} \omega$, where $\omega$ is a 
unit Gaussian random variable and $c_0$ a constant.
The crossover to pure diffusion occurs when considering ${\sf x} \propto \sqrt{4 D t}$,
in which case the fluctuations are Gaussian of magnitude
 proportional to $(\frac{r_0^2}{4 D t})^{1/4}$. A consistency check is that it corresponds to 
the lower edge of the validity of \eqref{approx}, i.e. ${\sf x}=x_0$ since one 
estimates $x_0=D/u = \frac{D t}{{\sf x}}$. The main message is that above
this scale, the physics of the KPZ equation sets in. 
\\

These arguments, although physically meaningful, are quite heuristic. 
The first step is the neglect of the terms $\partial_x (\xi P)$ above a small scale
$x_0$, $t_0$. The second is extending the results at fixed $u$, to smaller
values of $u={\sf x}/t$. From the results on the Beta polymer 
\cite{barraquand2017random,usBeta} we know that
the PDF, $P$, exhibits additional local fluctuations, with a Gamma distribution,
not present in the CDF, $P_>$, and this distinction does not seem
to be captured by the above arguments.

It would thus be very useful to obtain more controlled results. In this paper we 
provide some examples of exactly solvable models for which such controlled results can be obtained.
They confirm the above predictions and allow to refine the subleading error terms.

\subsection*{Quantum mechanical formulation} To close this section, and tie up with the following,
let us make some remarks on the analogy with quantum mechanics
which was very useful for the study of the KPZ equation and directed polymer problem
\cite{kardareplica}. Consider the model II defined above in \eqref{langevin},\eqref{Model2}, i.e. with a Gaussian random environment delta correlated 
in time and with a smooth short range correlation in space on scale $r_c$,
$\mathbb E\left[{\xi(x,t)  \xi(x',t')}\right] = r_0 D  \, \delta_{r_c}(x-x') \delta(t-t')$. The probability
$Q(x,t)= \mathbb{E}(\mathds{1}_{x(0)>0} | x(-t)=x)$ satisfies the backward Kolmogorov equation
\be
\partial_t Q(x,t)= D \partial_x^2 Q(x,t) + \xi(x,t) \partial_x Q(x,t),
\ee 
with initial condition $Q(x,0)=\mathds{1}_{x>0}$, and we use Ito prescription. One easily shows that the joint moments
\be
{\cal Q}_n(\vec x,t)  = \mathbb{E}\left[  \prod_{i=1}^n Q(x_i,t)\right]
\ee
satisfy 
\be \label{Hn} 
\partial_t {\cal Q}_n = - H_n {\cal Q}_n  , \quad  \quad
H_n = - D \sum_{i=1}^n \partial_{x_i}^2  - \frac{\sigma}{2} \sum_{1 \leq i \neq j \leq n} \delta_{r_c}(x_i-x_j) 
\partial_{x_i} \partial_{x_j} .
\ee
Here the operator $-H_n$ is the generator of the diffusion of $n$ particles in the same environment.
It can be interpreted as a quantum model of interacting bosons of
Hamiltonian $H_n$, but a different model from the standard Bose gas obtained to describe 
the (exponential) moments of the KPZ equation. 
Some aspects of the relation will be discussed in Appendix. 
A delta function version of this model will be solved below.

\section{An exactly solvable continuous model}
\label{sec:integrablediffusions}

The  model II defined by \eqref{langevin}  can be seen as a model of diffusion in random environment $\xi$. 
In this section, we study an exactly solvable model of diffusion in random environment introduced
by Le Jan and Raimond \cite{le2004sticky, le2004flows} and further studied in  \cite{le2004products, Howitt2009consistent, schertzer2015brownian, barraquand2019large}. This model cannot be defined by a stochastic differential equation  as in \eqref{langevin}, but rather using stochastic flows and the theory of sticky Brownian motions. We first review some general properties of sticky Brownian motions, and in a second stage consider the integrable model for which one can show convergence to
the KPZ equation in the moderate deviation regime.

\subsection{Sticky Brownian motions} 
\label{sec:stickyBM}
\begin{figure}[h]
\begin{center}
	\includegraphics[width=8cm]{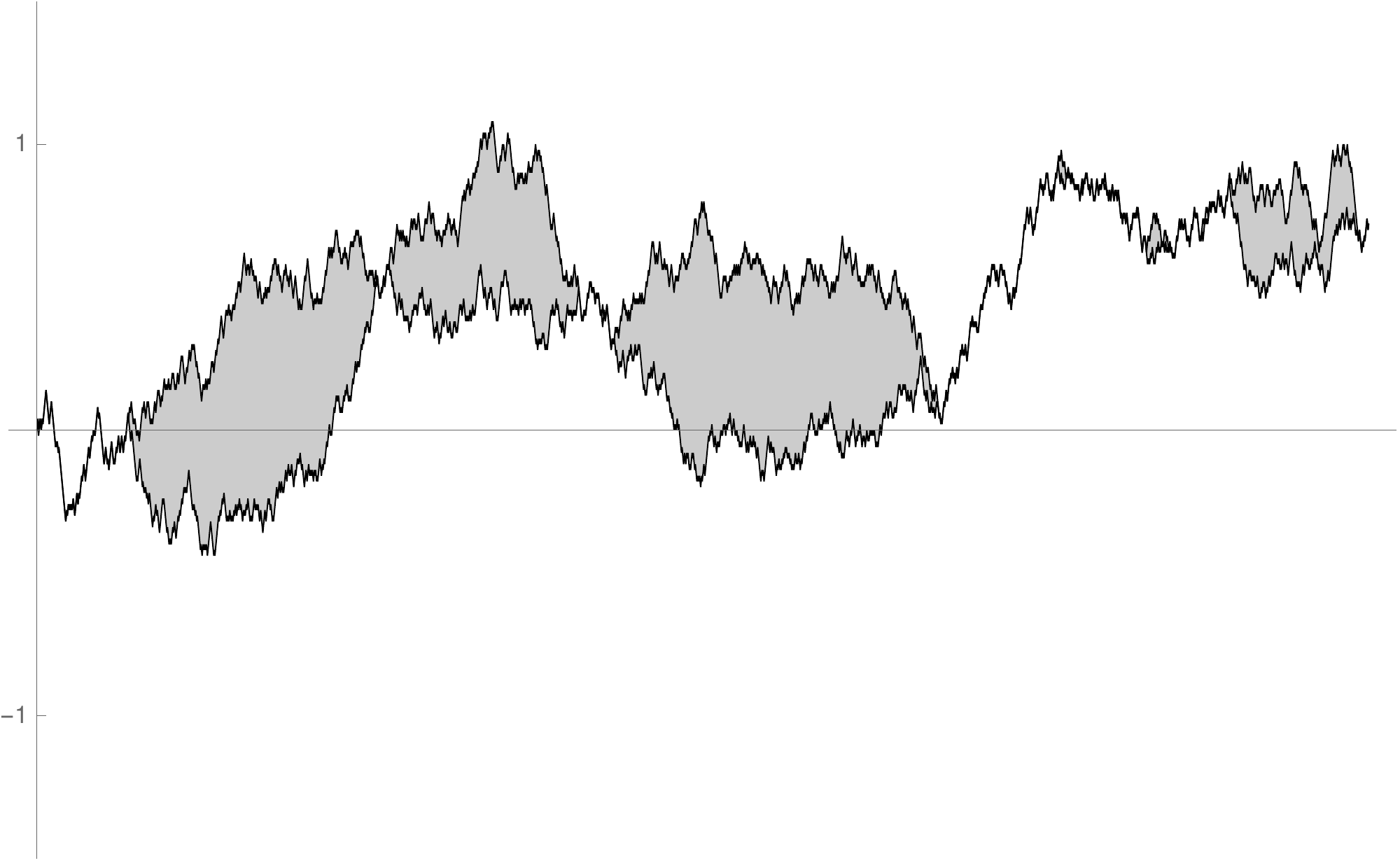}
	\includegraphics[width=8cm]{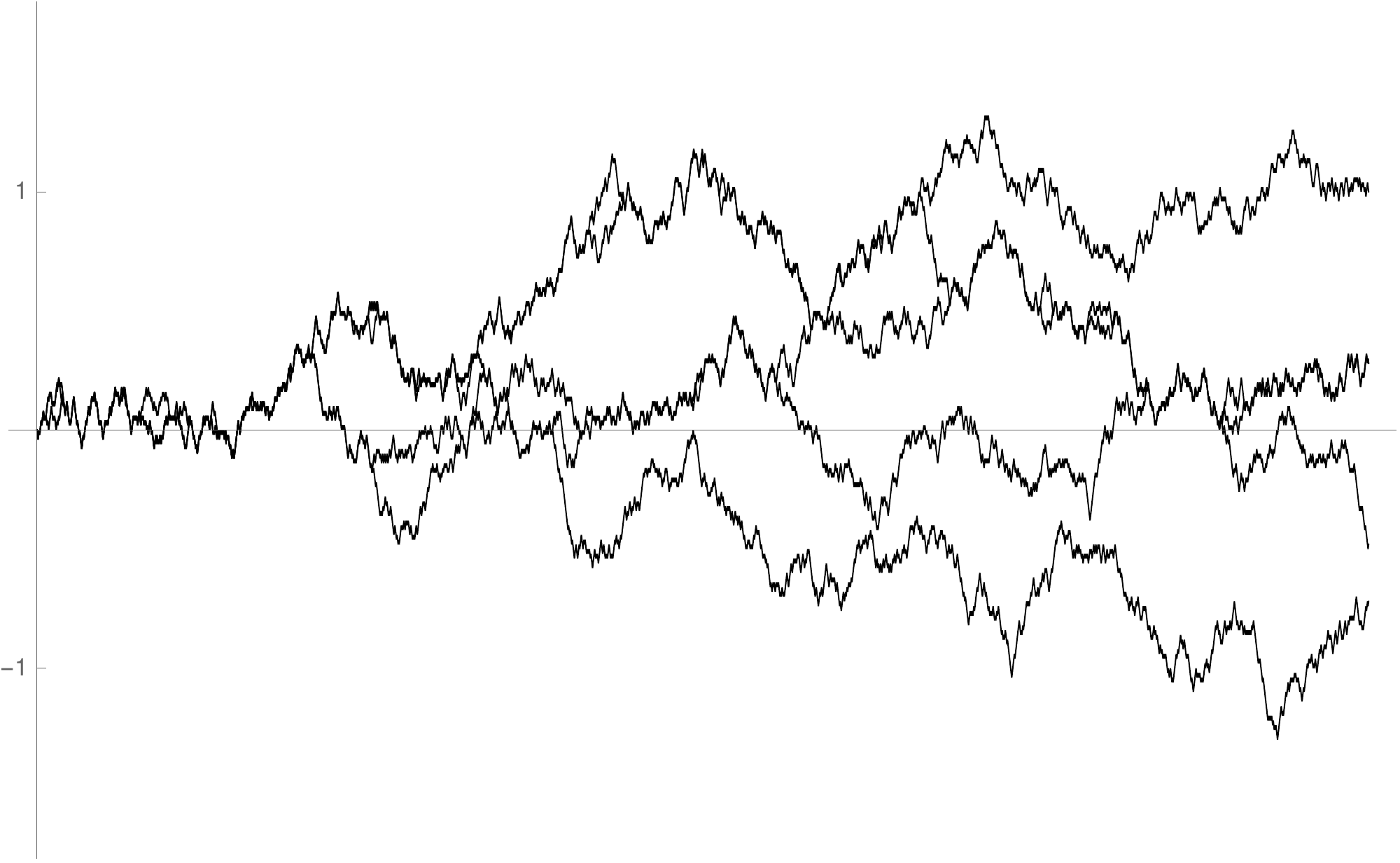}
\end{center}
\caption{Left: Simulation of $2$ sticky Brownian motions $x_1(t)$ and $x_2(t)$ starting from $x=0$ at $t=0$ up to time $t=1$ (taken from \cite{barraquand2019large}). Right: Simulation of $5$ sticky Brownian motions, note that for most of the time, at least two trajectories are stuck together. Both simulations were performed using a discretization of $n$-tuples of sticky Brownian motions (here $n=2$ or $5$) as $n$ random walks in the same space-time iid random environment (see more details in Section \ref{sec:convergencediscretetocontinuous} about this discretization).}
\label{fig:stickyBM}
\end{figure}

The limit of the model II (see \eqref{Model2}) as $r_c\to 0$ is mathematically quite subtle and was investigated in \cite{gawkedzki2004sticky, warren2015sticky}. Motivated by the study of Kraichnan's model \cite{kraichnan1968small} of passive scalar in turbulence (see also \cite{gawedzki1995anomalous, gawedzki1996university, bernard1998slow, gawedzki2000phase}) Gawedzki-Horvai \cite{gawkedzki2004sticky} showed that two independent diffusions $x_1(t), x_2(t)$  following the Langevin equation \eqref{langevin} with the same drift $\xi$ converge as $r_c$ goes to $0$ to a pair of so called sticky Brownian motions (the sticky interaction was first introduced in \cite{feller1952parabolic}). This means that $x_1(t)$ and $x_2(t)$ behave as standard Brownian motions when $x_1(t)\neq x_2(t)$, however, the two trajectories stick together in such a way that the Lebesgue measure of coincidence times between the two trajectories has a positive expectation (see Figure \ref{fig:stickyBM}, left). 

Warren \cite{warren2015sticky} further proved that $n$ independent diffusions $x_1(t), \dots, x_n(t)$  following \eqref{langevin} with the same drift $\xi$ converge as $r_c$ goes to $0$ to a $n$-tuple of sticky Brownian motions. More precisely \cite{warren2015sticky}  considered $n$  diffusions with generator
\be
\frac{r_c^2}{2} \sum_{j=1}^n \partial_{x_j}^2 + \frac{1}{2}\sum_{1 \leq i,j \leq n} \psi\left(\frac{x_i-x_j}{r_c}\right) \partial_{x_i} \partial_{x_j}
\label{eq:Warrendiffusion} 
\ee
where $\psi$ is a smooth function which decays at $\pm \infty$ and $\psi(0)=1$. The way these Brownian motions stick together in the $r_c \to 0$ limit, depends on a family of parameters $\theta(k,l)$ (which measure the rate at which $k+l$ paths stuck together will split into groups of $k$ and $l$ paths). See Figure \ref{fig:stickyBM} for an example of trajectory of $5$ sticky Brownian motions. The values of these $\theta(k,l)$ were explicitly computed in \cite{warren2015sticky} and  depend on $\psi''(0)$.

In general, families of sticky Brownian motions, can be considered from different points of view:
\begin{enumerate}
	\item As limits of discrete random walks in random environment: consider a random walk on the integers $\mathcal X_{T}$ indexed by $T\in \mathbb Z_{\geqslant 0}$, jumping by $\pm 1$ to one of  the nearest neighbour, with a probability $w_{X,T}$ (resp.  $1-w_{X,T}$) to jump from $X$ to $X+1$ (resp. $X-1$) between times $T$ and $T+1$, where the variables $w_{X,T}$ are iid. If $\mu^{\epsilon}(dw)$ denotes the law of the $w_{X,T}$ and $\epsilon^{-1}w(1-w)\mu^{\epsilon}(dw) \Rightarrow \nu(dw)$ for some finite measure $\nu$ on $(0,1)$  as $\epsilon \to 0^+$, then 
	 $x^{\epsilon}(t) = \epsilon \mathcal X_{\epsilon^{-2}t}$ converges 
to  a diffusion $x(t)$ in a random environment. Likewise, several random walks $\mathcal X^1_{T},\mathcal X^2_{T} \dots$ (sampled independently but using the same transition probabilities $w_{X,T}$) converge to diffusions $x_1(t),x_2(t),\dots$ which can be seen as independent diffusions in the same environment.  These diffusions will have the same joint distribution as sticky Brownian motions where the splitting rates $\theta(k,l)$ are related to the measure $\nu$ via 
	\begin{equation}
	\theta(k,l) = \int_0^1 w^{k-1}(1-w)^{l-1}\nu(dw).
	\end{equation}
	This convergence was first studied in \cite{le2004products} in a specific case, and then more generally in \cite{Howitt2009consistent}.
	\item As independent diffusions in a random environment. The environment was constructed explicitly  \cite{schertzer2014stochastic} (see also the review \cite{schertzer2015brownian}) using results on  the so-called Brownian web \cite{arratia1979coalescing, toth1998true} and Brownian net \cite{sun2008brownian}. The diffusions follow paths of the Brownian web/net, and at special points of the Brownian web/net (classified in \cite{toth1998true, newman2010marking, fontes2004brownian, schertzer2009special}), trajectories may branch according to a random variable of law $\nu$ (potentially renormalized to be a probability measure),  we refer to \cite{schertzer2015brownian} for details. 
	\item As a diffusion process characterized by a martingale problem, see  \cite{schertzer2015brownian, Howitt2009consistent}. 	
\end{enumerate} 

Sticky Brownian motions fit in the theory of stochastic flows of kernels \cite{le2004flows, le2002integration}. A stochastic flow of kernels is a family of random probability measures $
\mathsf{K}_{st}(x,dy)$ which represent the probability for a diffusion starting at $x$ at time $s$ to arrive in $[y,y+dy]$ at time $t$. For any family of sticky Brownian motions characterized by a given sequence of 
parameters $\theta(k,l)$ there exist a stochastic flow of kernels such that $n$ sticky Brownian motions corresponds to $n$ diffusions having transition probabilities equal to the product
$\mathbb E\left[\prod_{i=1}^n \mathsf{K}_{st}(x_i,dy_i)\right]$.

\subsection{Integrable model} 
\label{sec:LeJanRaimondflow}

In this section, we consider families of sticky Brownian motions depending on a single parameter $\lambda>0$, and the corresponding diffusions in random environment, in the special case where 
\begin{equation}
\theta(k,l) = \frac{\lambda}{2}  B(k,l) = \frac{\lambda}{2} \frac{\Gamma(k)\Gamma(l)}{\Gamma(k+l)}.
\label{eq:choicethetaintegrable}
\end{equation} 
This family of sticky Brownian motions was considered by \cite{le2004sticky, le2004products} 
in the context of stochastic flows of kernels.
It was shown in  \cite{barraquand2019large} that this model is exactly solvable. 
It can also be formally associated with the generator
\be \label{gen} 
  \sum_{i=1}^n \frac 1 2 \partial_{x_i}^2  + \frac{1}{\lambda} \sum_{1 \leq i < j \leq n} \delta(x_i-x_j) 
\partial_{x_i} \partial_{x_j}.
\ee

\begin{remark} Although the function $\delta_{r_c}$ in \eqref{Hn} converges in the limit $r_c\to0$ to a delta function $\delta$ as in \eqref{gen},  this integrable model is different from the $r_c\to 0$ limit obtained by Warren in \cite{warren2015sticky}. Indeed, \cite{warren2015sticky} obtains in the limit sticky Brownian motions characterized by rates $\theta(k,l)$ which are explicit but depend on the second derivative of the function $\psi$ in \eqref{eq:Warrendiffusion} and are clearly different from the choice of $\theta(k,l)$ given in \eqref{eq:choicethetaintegrable} which makes the model integrable.
\end{remark} 
Let $\mathsf K_{-t,0}(x, [0,+\infty))$ be the probability for a diffusion in random environment (in the special case \eqref{eq:choicethetaintegrable}) starting at $x$ at time $-t$ to end in $[0,+\infty)$ at time $0$. 
Let us define
\begin{equation}
{\cal Q}^{\lambda}_k(\vec x,t)= \mathbb E \Big[\mathsf K_{-t, 0}(x_1, [0, +\infty))\dots \mathsf K_{-t, 0}(x_k, [0, +\infty)) \Big].
\label{eq:defQlambda}
\end{equation}
For $x_1\geqslant \dots \geqslant x_k$, and  $t\geq 0$, we have \cite[Proposition 1.20]{barraquand2019large}
\begin{equation}
{\cal Q}^{\lambda}_k(\vec x,t) =  \\ \int_{\alpha_1+\I\mathbb R} \frac{\mathrm d w_1}{2\I\pi} \dots \int_{\alpha_k+\I \mathbb R}  \frac{\mathrm d w_k}{2\I \pi} \prod_{1\leqslant A<B \leqslant k} \frac{w_{B}-w_A}{w_B-w_A-w_Aw_B} \prod_{j=1}^k \exp\left( \lambda^2 w_j^2t/2 +\lambda x_jw_j \right)\frac{1}{w_j},
\label{eq:formulaforPhit}
\end{equation}
where for $i<j$, $ 0<\alpha_i < \frac{\alpha_{j}}{1+\alpha_{j}}$. 
 Equivalently, we may use the change of variables $z_i=1/w_i$ in \eqref{eq:formulaforPhit} and write 
\begin{equation}
{\cal Q}^{\lambda}_k(\vec x,t) =  \\ \int_{r_1+\I\mathbb R} \frac{\mathrm d z_1}{2\I\pi} \dots \int_{r_k+\I \mathbb R}  \frac{\mathrm d z_k}{2\I \pi} \prod_{1\leqslant A<B \leqslant k} \frac{z_{A}-z_B}{z_A-z_B-1} \prod_{j=1}^k \exp\left(\frac{t \lambda^2}{2z_j^2} +\frac{\lambda x_j}{z_j} \right)\frac{1}{z_j},
\label{eq:formulaforPhitbis}
\end{equation}
where for $i<j$, $ r_i>r_j+1$.

One may observe \cite[Section 6]{barraquand2019large} that ${\cal Q}^{\lambda}_k(\vec x,t)$ is a solution $u(\vec x,t)$ of the following heat equation on 
$\lbrace \vec x \in \mathbb{R}^k : x_1\geqslant \dots\geqslant x_k \rbrace$  subject to boundary conditions
\begin{equation} \label{eq:freeevolutionwithboundary}
\begin{cases} \partial_t u(\vec x,t)= \tfrac{1}{2}\Delta u(\vec x,t), \quad t\geq 0, \vec x\in \mathbb R^k,\\
(\partial_{x_i} \partial_{x_{i+1}} +  \lambda (\partial_{x_i}-\partial_{x_{i+1}})) u(\vec x,t)|_{x_i=x_{i+1}}=0. 
\end{cases}
\end{equation}

It is natural to associate to \eqref{eq:freeevolutionwithboundary} the following equation on $\mathbb R^k$ with point interactions, which involves the generator \eqref{gen}
\begin{equation}
\partial_t v(t, \vec x)= \tfrac{1}{2}\Delta v(t,\vec x) + \frac{1}{2 \lambda} \sum_{i \neq j}\delta(x_i-x_j) \partial_{x_i} \partial_{x_j}v(t, \vec x).  
\label{eq:trueevolution}
\end{equation}
so that $v=u$ on the set $\lbrace \vec x \in \mathbb{R}^k : x_1\geqslant \dots\geqslant x_k \rbrace $.

We now state two important properties which will imply the convergence to the KPZ equation
in the moderate deviation regime. First, the exact formula \eqref{eq:formulaforPhit} obeys the scaling property
\begin{equation}
{\cal Q}_k^{\lambda}(\vec x,t) = {\cal Q}_k^{1}(\lambda \vec x,\lambda^2 t).
\label{eq:scalingproperty}
\end{equation}
Second it was shown in \cite[Proposition 6.3]{barraquand2019large} that the moments of
the stochastic flow of kernels converge, in the weak disorder and large deviation regime, to
the moments of the SHE
\begin{equation}
\lim_{\lambda\to+\infty} \frac{{\cal Q}^{\lambda}_k(-\lambda^2t-\lambda \vec x,\lambda^2t)}{\prod_{i=1}^k C_{\lambda,t,x_i}} = \mathbb E\left[\prod_{i=1}^k Z(x_i,t/2)\right],
\end{equation}
where $Z(x,t)$ is the solution of the SHE \eqref{she} with droplet initial data, $Z(x,t=0)=\delta(x)$
and
\begin{equation}
C_{\lambda, t, x} = \frac{1}{\lambda} \exp\left( -\frac{\lambda^2t}{2}-\lambda x \right).
\end{equation}
is a normalization factor.

Using the scaling \eqref{eq:scalingproperty} we can convert asymptotics in the weak noise
large deviation regime into asymptotics in the moderate deviation regime with fixed amplitude of the 
noise. Denoting $L=\lambda^4$ we obtain
\begin{equation}
\lim_{L\to+\infty} \frac{{\cal Q}_k^{1}(- L^{3/4}t-L^{1/2} \vec x, L t)}{\prod_{i=1}^k C'_{L,t,x_i}} = \mathbb E\left[\prod_{i=1}^k Z(x_i,t/2)\right],
\label{eq:convergenceofmomentsmoderatedeviations}
\end{equation}
where 
\begin{equation}
C'_{L, t, x} = \frac{1}{L^{1/4}} \exp\left( -\frac{L^{1/2}t}{2}-L^{1/4} x \right).
\end{equation}

Equivalently, this may be interpreted as the following. For $\tilde x>0$ and $y\in \mathbb R$ and $\lambda=1$, 
\begin{equation}
\log {\mathsf P}\left(x(0) \geqslant 0 \big\vert x(-t) = - (\tilde x t^{3/4} + t^{1/2} y) \right)  + \frac{1}{2} \tilde x^2t^{1/2}+ \tilde x t^{1/4} y + \frac{1}{4}\log(t) -  \log(\tilde x)
\xRightarrow[t \to +\infty]{} h^{\rm KPZ}(y \tilde x^{2}, \tilde x^4/2),
\label{eq:convergenceexplicit}
\end{equation}
in the sense that the exponential moments of the left-hand-side converge to the exponential moments of the right hand side. We expect the convergence to be true in terms of convergence in distribution, and not only for fixed $y, \tilde x$, but as well as a process indexed by  $y, \tilde x\in \mathbb R \times\mathbb R_{>0}$.

\begin{remark}
In general, the convergence of moments does not imply the weak convergence of probability measures. In the present case, the moments of $Z(x, t)$ grow too fast to even determine uniquely the distribution. Thus, the convergence of moments \eqref{eq:convergenceofmomentsmoderatedeviations} does not imply distributional convergence, but this a very strong indication of it. Proving the  convergence in distribution remains a mathematical open problem.
 \label{rem:convergenceofmomentsvsweak}
\end{remark}

\begin{remark}
	We can compare the exact result \eqref{eq:convergenceexplicit} with the prediction \eqref{predict}.
	To compare them, we must set $v_0=0$, $D=1/2$ and $r_0=2$ in \eqref{predict} so that $\sigma=1$. 
	One then sees that 
	$\tilde {\sf x}^4$ in \eqref{predict} becomes identical to $\tilde x^4/2$ in \eqref{eq:convergenceexplicit}
	(note that $x,t$ are the same coordinates in both models). The KPZ term is thus identical. 
	The diffusion term $-x^2/(4 D t)$ in \eqref{predict} also equals $\frac{1}{2} \tilde x^2 t^{1/2}$
	in \eqref{eq:convergenceexplicit}. Note that the comparison is made setting $y=0$, since
	\eqref{eq:convergenceexplicit} contains
	additional finer information about the spatial dependence on scales $t^{1/2}$. 
	Finally the log term is absent in \eqref{predict}.
	
	In addition, in the large deviation regime, $x = u t$ but for small values of $u$, one
	can also match the prediction \eqref{res2} with the exact result  in \cite[Theorem 1.13 and
	1.15]{barraquand2019large}. Indeed the rate function obtained there	has the following expansion at small $u$ (where $u=x/t$ in both cases), $J(u)= \frac{u^2}{2} + \frac{u^4}{24} + O(x^6)$, and the function $\lambda(u)$ denoted $\sigma(u)$ there \footnote{There is a typo in the expression given for 
$\sigma(u)$ in \cite{barraquand2019large}.  In \cite[Equation 
(12)]{barraquand2019large}, in the expression for $\sigma$, the factor 
$1/2$ should be
$1/2^{1/3}$, so that $2 \sigma^3=h'''(\theta)$.} 
	 behaves as 
	$\lambda(u) = 2^{-1/3} u^{4/3}$. Setting again $D=1/2$ and $r_0=2$
	in \eqref{res2} reproduces this result.

\end{remark}

\section{An exactly solvable discrete model}
\label{sec:RWRE} 

In this section, we study a discrete analogue of the model considered in Section \ref{sec:LeJanRaimondflow}, for which a  similar convergence towards the KPZ equation in the moderate deviation regime can be proved. 

\subsection{Random walk in Beta distributed random environment} 
Let $\alpha, \beta>0$ be two parameters. Let $(B_{T,X})$ be a family of iid random variables in $(0,1)$, indexed by $T\in \mathbb Z_{\geqslant 0}$ and $ X\in \mathbb Z$, marginally distributed according to the Beta distribution with parameters $\alpha, \beta$, i.e. with density 
\begin{equation}
\mathbb P(B\in dx) = \frac{\Gamma(\alpha+\beta)}{\Gamma(\alpha)\Gamma(\beta)} x^{\alpha-1}(1-x)^{\beta-1}.
\end{equation} 
For a fixed $X\in \mathbb Z$, consider the discrete time nearest neighbour random walk in random environment  $(\mathcal X_T^{X})_{T\geqslant 0}$ in $\mathbb Z$ (called Beta RWRE below), starting from $\mathcal X_0^X=X$ (we will sometimes omit the superscript when the starting point is clear from the context or does not matter), with random transition probabilities 
\begin{equation}
\mathsf P(\mathcal X_{T+1}=X+1\vert  \mathcal X_{T}=X) = B_{T,X}, \;\;\mathsf P(\mathcal X_{T+1}=X-1\vert \mathcal X_{T}=X) = 1-B_{T,X}.
\end{equation}
We use the symbol $\mathsf P$ to denote the probability measure of random walks conditionally on the environment (quenched), while $\mathbb P$ denotes the probability measure of the environment, and $\mathbb E$ is the expectation associated to $\mathbb P$ as above. 
Let us define  
\begin{equation}
{\cal Q}^{\rm RW}_k(\vec X,T)  = \mathbb E \left[ \prod_{i=1}^k \mathsf P(\mathcal X_T^{X_i}\geqslant 0) \right]. 
\label{eq:defQRW}
\end{equation}
The time evolution of the function ${\cal Q}^{\rm RW}_k$ can be related to a difference operator acting on functions of the variable $\vec X$, which can be diagonalized  using coordinate Bethe ansatz \cite[Section 3]{barraquand2017random}. For $X_1\geqslant X_2 \geqslant \dots \geqslant X_k$, one has  \cite[Proposition 3.4]{barraquand2017random}
\begin{equation}
\label{eq:betarwremoment} 
{\cal Q}^{\rm RW}_k(\vec X,T) = \\  \int_{\gamma_1} \frac{dz_1}{2\mathbf i \pi}  \dots \int_{\gamma_k} \frac{dz_k}{2\mathbf i \pi}  \prod_{A<B} \frac{z_A-z_B}{z_A-z_B-1} \prod_{j=1}^k \left(\frac{ \alpha+\beta+z_j}{z_j} \right)^{\frac{T+X_i}{2}+1} \left( \frac{ \alpha +z_j}{\alpha+\beta+z_j} \right)^T \frac{1}{z_j+\alpha+\beta},
\end{equation}
where the contours $\gamma_i$ are positively oriented closed curves around $0$ such that for $i<j$, $\gamma_i$ contains $\gamma_j+1$ and all contours exclude $-\alpha-\beta$. \cite[Proposition 3.4]{barraquand2017random} deals with a variant of the model called Beta polymer therein, the translation of the result in terms of the Beta RWRE defined above was given in \cite{barraquand2017random} and more explicitly in \cite[Proposition 6.2]{barraquand2019large}.

\subsection{Convergence to the continuous model}
\label{sec:convergencediscretetocontinuous}

As we have already mentioned in Section \ref{sec:stickyBM}, families of  discrete random walks in space-time iid random environment converge to families of sticky Brownian motions. Consider a simple random walk on the integers $\mathcal X_{T}$ indexed by $T\in \mathbb Z_{\geqslant 0}$, jumping by $\pm 1$ to one of the nearest neighbour, 
\begin{equation}
\mathsf P(\mathcal X_{T+1}=X+1\vert  \mathcal X_{T}=X) = w_{T,X}, \;\;\mathsf P(\mathcal X_{T+1}=X-1\vert \mathcal X_{T}=X) = 1-w_{T,X}.
\end{equation}
 where the variables $w_{X,T}$ are iid and are distributed according to a probability measure  $\mu^{\epsilon}(dw)$. 

Assume that $\epsilon^{-1}w(1-w)\mu^{\epsilon}(dw) \Rightarrow \nu(dw)$ for some finite measure $\nu$ on $(0,1)$. Then, it was shown in \cite{Howitt2009consistent} (see also \cite[Section 5]{schertzer2015brownian}) that   $x^{\epsilon}(t) = \epsilon \mathcal X_{\epsilon^{-2}t}$ converges to  a diffusion $x(t)$ in a random environment  as $\epsilon \to 0^+$. Several diffusions $x_1(t),x_2(t),\dots$ in the same environment will have the same joint distribution as sticky Brownian motions where the splitting rates $\theta(k,l)$ are related to the measure $\nu$ via 
\begin{equation}
\theta(k,l) = \int_0^1 w^{k-1}(1-w)^{l-1}\nu(dw).
\end{equation}
This convergence was first studied in \cite{le2004products} in a specific case (Beta RWRE on the torus), and then proved for  more generally random walks in random environment in \cite{Howitt2009consistent}.

In the case of the Beta RWRE, the limiting disorder measure $\nu$ is $\nu(dx) = \frac{\lambda}{2} dx$, 
 so that the  Beta RWRE  model converges to the model studied in Section \ref{sec:LeJanRaimondflow} when the parameters $\alpha, \beta$ of the Beta RWRE are scaled as $\alpha=\beta=\lambda \epsilon$, and paths are rescaled diffusively as above. In particular, recalling the notations ${\cal Q}^{\rm RW}_k(\vec X,T)$ from \eqref{eq:defQRW} and ${\cal Q}^{\lambda}_k(\vec x,t)$  from \eqref{eq:defQlambda}, we have the convergence  
\begin{equation}
{\cal Q}^{\rm RW}_k(\epsilon^{-1}\vec x,\epsilon^{-2} t) \xrightarrow[\epsilon\to 0]{} {\cal Q}^{\lambda}_k(\vec x,t).
\end{equation} 

\subsection{Convergence to the KPZ equation}
Note that for $\vec X=(-X, -X, \dots , -X)$, the function ${\cal Q}^{\rm RW}_k$ corresponds to the $k$th moment of the CDF of a random walk in random environment starting from $0$, that is 
\begin{equation}
{\cal Q}^{\rm RW}_k(\vec X,T)  = \mathbb E \left[\mathsf P(\mathcal X_T^0\geqslant X)^k\right], 
\end{equation}
so that the knowledge of the functions ${\cal Q}^{\rm RW}_k(\vec X,T) $ fully characterizes the probability distribution of the random variable $\mathsf P(\mathcal X_T^0\geqslant X)$. 

Let $v_0=(\alpha-\beta)/(\alpha+\beta)$ be the average drift of the random walk $\mathcal X_T$. For $T=tL$ and $X=x\sqrt{L}$, it can be shown that $\mathsf P\left(\frac{\mathcal X_T- v_0 T}{\sqrt{1- v_0^2}} \geqslant X\right)$ converges as $L$ goes to infinity to the deterministic limit  $\int_{x/\sqrt{t}}^{+\infty} dy e^{-y^2/2}/\sqrt{2\pi}$ \cite{rassoul2009almost}.
This is the Gaussian regime of typical events.

When, however, one considers large deviation events, i.e. for $T=L$ and $X=x L$,  then 
\be
 \log \mathsf P(\mathcal X_T\geqslant X) \simeq -L I(x) + \sigma(x)L^{1/3} \chi_{\rm GUE},
 \ee
where $\chi_{\rm GUE}$ denotes a random variable following the Tracy-Widom GUE distribution, for some explicit functions $I(x), \sigma(x)$ \cite{barraquand2017random, usBeta}
which depend on $\alpha, \beta$.

In this paper, we are interested in the intermediate regime where $X \propto L^{3/4}$. More precisely, let us consider the scalings 
\begin{eqnarray}
X &=& -2 \alpha^2 t L^{3/4} - \alpha x L^{1/2} \label{eq:scalingX},\\
T&=& 2 \alpha^2 t L \label{eq:scalingT},\\
\beta &=& \alpha - V/\sqrt{L}, \label{eq:scalingbeta}\\
z &\to& z - \alpha + \alpha L^{1/4}. \label{eq:scalingz}
\end{eqnarray}
Then, for $x_1\leqslant \dots \leqslant x_k$, 
\begin{eqnarray}
\lim_{L\to\infty} \frac{{\cal Q}^{\rm RW}_k(\vec X,T) }{\prod_{i=1}^k C''_{L,t,x_i}} &=& 
\int_{r_1+\I\mathbb R} \frac{dz_1}{2\mathbf i \pi}  \dots \int_{r_k+\I\mathbb R} \frac{dz_k}{2\mathbf i \pi}  \prod_{A<B} \frac{z_A-z_B}{z_A-z_B-1} \prod_{j=1}^k e^{tz_i^2 +(x_i-V t) z_i},\label{eq:mixedmomentsSHE}\\
&=&  \mathbb E\left[\prod_{i=1}^k Z(x_i-V t,t)\right].  \label{eq:mixedmomentsSHEbis}
\end{eqnarray}
where the renormalization factor $C''_{L,t,x}$ is given by 
\begin{equation}
C''_{L,t,x} = \frac{1}{\alpha L^{1/4}} \exp\left(- \alpha^2 t \left(\sqrt{L}+\tfrac{1}{6}\right) - \alpha L^{1/4} (x - V t)\right)
\end{equation}
and the contours are vertical lines oriented from bottom to top with real parts $r_i$ such that for $i<j$, $r_i>r_j+1$. Equation \eqref{eq:mixedmomentsSHE} is readily obtained by plugging the scalings \eqref{eq:scalingX}, \eqref{eq:scalingT}, \eqref{eq:scalingbeta}  and the change of variables \eqref{eq:scalingz}  into the integral formula \eqref{eq:betarwremoment}. The limit can be mathematically justified by dominated convergence, we refer to \cite{borodin2014macdonald, corwin2017kardar, barraquand2019large} where very similar limits were considered. The equality between the mixed moments of the multiplicative SHE in \eqref{eq:mixedmomentsSHEbis}  and the integral formula \eqref{eq:mixedmomentsSHE} was shown in  \cite[Proposition 6.2.3]{borodin2014macdonald} (see also \cite{ghosal2018moments}), modulo some factors of $2$ accounting for the difference of conventions between the physics and the mathematics literature.

Hence, we have shown that under the scalings \eqref{eq:scalingX}, \eqref{eq:scalingT}, \eqref{eq:scalingbeta},  the random variable $\mathsf P(\mathcal X_T\geqslant -X)$, when appropriately rescaled, converges to  $Z(x-V t, t)$, in the sense that all moments converge.  Again, convergence in distribution is an open problem from the mathematical point of view (see Remark \ref{rem:convergenceofmomentsvsweak}),  although it was rigorously proved under a different scaling in \cite{corwin2017kardar} using chaos series expansions. Note also that in order to formulate the convergence of mixed moments in space and time, one a priori has to vary the starting points of the Beta RWRE, not the ending points, and reverse time.  

Equivalently, this may be interpreted as the following. For $\tilde x>0$ and $y\in \R$,
\begin{multline}
\log {\mathsf P}\left(\mathcal X_{0} \geqslant 0 \big\vert \mathcal X_{-T} = - (
 \tilde x T^{3/4} + y T^{1/2} + \frac{V \tilde x^2}{2\alpha} T^{1/2} )  \right) + \frac{\tilde x^2 T^{1/2}}{2}+ \tilde x y T^{1/4} + \frac{1}{4}\log(T) -  \log(\tilde x/\alpha) +\frac{\tilde x^4}{12}  \\
\xRightarrow[T \to +\infty]{} h^{\rm KPZ}\left(\frac{y \tilde x^{2}}{\alpha}, \frac{\tilde x^4}{2\alpha^2}\right),
\label{eq:convergencenotweak}
\end{multline}
where on the l.h.s. the Beta RWRE has a fixed parameter $\alpha$ whereas $\beta$ is scaled
so that $\alpha- \beta=V \tilde x^2 T^{-1/2}$. This induces as drift $\mathbb{E}[\mathcal X_{T}- \mathcal X_{0}] = \frac{V}{2 \alpha}  
\tilde x^2 T^{1/2}$, which accounts for the same factor inside the l.h.s. Hence $\tilde x T^{3/4}+y T^{1/2}$ represent the deviations with respect to the typical walk. Again, the convergence \eqref{eq:convergencenotweak} means that the exponential moments of the l.h.s. converge to those of the r.h.s., but we expect that the convergence holds 

\begin{remark}
	The convergence of quenched transition probabilities in the Beta RWRE model to the KPZ equation was already proven in \cite{corwin2017kardar}, though this was under a different scaling. Indeed, models in the KPZ universality class depending on some tunable parameters generally converge to the KPZ equation under appropriate scaling. Two specific scalings are often considered in the literature, a scaling of weak noise and a scaling of weak asymmetry (weak non-linearity scaling). In the Beta RWRE model, there are two parameters $\alpha, \beta$ so that one can use both scalings (or a mixture of the two). The result of \cite{corwin2017kardar} corresponds to convergence to the KPZ equation under weak noise scaling, and this requires to consider the quenched probability of large deviation events. The scaling that we considered above corresponds to the convergence to the KPZ equation under weak asymmetry scaling. 
	\label{rem:CorwinGu}
\end{remark}

\section{Extrema of many diffusions}
\label{sec:maximum} 

A natural setting to observe in experiments the sample to sample fluctuations in the 
tail of the probability distribution $P_>(x,t)={\rm Prob}(x(t)>x)$ is to consider a large number $N \gg 1$ of independent
particle $x_i(t)$ in the same environment, and to measure the position of the rightmost particle 
$x_{\rm max}(t) = \max_{i=1,\dots N} x_i(t)$ (we focus here on one space dimension) 
As is well known, for $N$ particles
performing independent Brownian motions, the position of the maximum at a given time $t$
is distributed for $N \gg 1$ as
\be \label{gumb}
x_{\rm max} = \sqrt{4 D t} \left(\sqrt{\log N} + \frac{G - c_N}{2 \sqrt{\log N}}\right)  , \quad \quad {\rm Prob}(G<g) = e^{- e^{-g}}  , \quad \quad c_N = \frac{1}{2} \log(4 \pi \log N),
\ee 
where $G$ is a Gumbel random variable, describing the ``thermal'' noise experienced by the particles.
This is the usual Einstein description of the motion of a tracer molecule in a fluid, above the collision scale,
where $D$ is the molecular diffusion coefficient. Here however we are interested
in a more subtle effect, which arises by considering the fluid as a time-dependent
random environment. In that case one predicts sample to sample fluctuations
of $x_{\rm max}$ as we now discuss. Note that this is a non-trivial deviation from Einstein's
theory and arises because some tracer particles can take advantage of ``good spots'' in the space-time
history of the environment where they receive a strong bias to the right. Hence it requires a model with random local bias.
Whether it can be detected in experiments or not is as yet an
open question, but it is important to refine the predictions, and search for the
most favorable conditions (i.e. $N$ not astronomically large and the sample to sample
variance of detectable amplitude). We expect that the moderate deviation regime 
studied here is a good candidate for that.

This effect of the time-dependent environment was first predicted and studied rigorously in \cite{barraquand2017random} for the lattice model of the Beta random walk, and later in \cite{PLDTTDiffusion} for a continuum diffusion model, using the qualitative arguments recalled in Section 2.

Let us recall in simple terms the analysis, starting with the large deviation regime (for rigorous statements and derivations see \cite{barraquand2017random}) and then considering the moderate
deviation regime which is the focus of this paper. The standard result for $N\gg1 $ independent random variables gives that
\be
{\rm Prob}(x_{\rm max}<x) = e^{N \log (1- P_>(x,t)) } \simeq e^{ - N P_>(x,t)} = e^{- e^{\log N + \log P_>(x,t)}} 
\ee 
since $P_>(x,t)$ is small in the region where the argument of the double exponential 
is of order unity (i.e. one probes the tail of $P_>$). Since the Gumbel CDF is 
${\rm Prob}(G<g) = e^{- e^{-g}}$, the random variable $x_{\rm max}$ is distributed so that
\be \label{eq0} 
\log N + G  + \log P_>(x_{\rm max},t) = 0 .
\ee
When $P_>$ is the CDF of the simple Gaussian, it gives 
$\frac{x_{\rm max}^2}{4 D t} = \log N + G +\frac{1}{2} \log(D t/\pi x_{\rm max}^2)$, recovering \eqref{gumb}.
Note that \eqref{eq0} does not necessarily imply Gumbel fluctuations for $x_{\rm max}$
and remains valid in the two other domains of attraction of max stable distributions
(e.g. if $P_>(x,t) \propto x^{-\alpha}$ \eqref{eq0} leads to Frechet fluctuations for
$x_{\rm max}$ since $e^{G/\alpha}$ has Frechet distribution). The models considered here
however are in the Gumbel domain of attraction.
In presence of
the random environment $\log P_>$ additionally fluctuates from sample to sample, and
so does $x_{\rm max}$. One distinguishes two regimes.

\subsection{Large deviation regime: TW fluctuations}
The large deviation regime with $u=\frac{x}{t}$ fixed, which leads
to KPZ class fluctuations, is probed by considering both $\log N$ and $t$ large, with a fixed ratio $\gamma=\frac{1}{t} \log N$
\cite{barraquand2017random,PLDTTDiffusion}. This leads to a maximum growing
linearly with time, $x_{\rm max} \propto t$.
From \eqref{eq0} and \eqref{res2} one obtains 
\be
J(u_m) t = \log N + G  +  \lambda(u_m) t^{1/3} \chi_2 + o(t^{1/3})  ,\quad \quad u_m=\frac{x_{\rm max}}{t},
\ee
where the rate function $J(u)$ and $\lambda(u)$ depend on the model.
To leading order, the position of the maximum is determined by $J(\frac{x_{\rm max}^0}{t})=\gamma$,
hence $x_{\rm max}^0= J^{-1}(\gamma) t$. Denoting $x_{\rm max} = x_{\rm max}^0 + \delta x_{\rm max}$ and expanding
in powers of $\delta x_{\rm max}$ one finds
\be \label{tw0} 
x_{\rm max} \simeq J^{-1}(\gamma) t + \frac{1}{J'(J^{-1}(\gamma))} \left( \lambda\left(J^{-1}(\gamma)\right) t^{1/3} \chi_2 
+ o(t^{1/3}) + G  \right).
\ee 
For the continuum diffusion model II, using $J(u)$ and $\lambda(u)$ given in \eqref{res2}, 
the arguments in Section 2 predict, in the limit of small $\gamma$, precisely in an expansion in
$\frac{\gamma r_0^2}{D} \ll 1$
\be \label{tw} 
x_{\rm max} \simeq t \sqrt{4 D \gamma}  \left(1 - \frac{\gamma r_0^2}{96 D} + O\left(\left(\tfrac{\gamma r_0^2}{D}\right)^2\right) \right)
+ \frac{1}{2} r_0^{2/3} (4 D \gamma)^{1/6} t^{1/3} \chi_2 + o(t^{1/3}) + \sqrt{\frac{D}{\gamma}} G .
\ee
In Eqs. \eqref{tw0}, \eqref{tw} the term containing $G$ represents the Gumbel ``thermal'' fluctuations for
different particles in the same environment, while the term containing $\chi_2$, the TW random variable,
as well as its subleading correction denoted $o(t^{1/3})$ represents the sample to sample fluctuations. Note that in that regime the latter which grow as $t^{1/3}$ are larger than the former. 

\subsection{ Moderate deviation regime: KPZ equation fluctuations}
 We now consider the regime $\log N \propto \sqrt{t}$, more precisely we define the dimensionless parameter $g$ 
such that 
\be
\log N = g \frac{\sqrt{4 D t}}{r_0} ,
\ee 
and consider the limit where $N,t$ are large at fixed $g$. 
From \eqref{eq0} and \eqref{predict} one predicts that $x_{\rm max}$ is distributed such that
\be
\frac{x_{\rm max}^2}{4 D t} + O(\log t) = \log N + G +  h^{\rm KPZ}(0,\tilde x_{\rm max}^4)  ,\quad \quad 
\tilde x_{\rm max}= \frac{x_{\rm max}}{r_0 (\frac{4 D t}{r_0^2})^{3/4}},
\ee 
where the $O(\log t)$ term is deterministic and cannot be predicted by the 
arguments in Section \ref{sec:physics} (it includes the effect of considering
the CDF rather than the PDF). Hence the sample to sample fluctuations will be controlled by the droplet solution, $h^{\rm KPZ}$, 
of the KPZ equation defined in \eqref{KPZh}, at an "effective time" equal to $\tilde x_{\rm max}^4$. More precisely,
expanding around the leading behavior $x_{\rm max} = x_{\rm max}^0=\sqrt{4 D t \log N}$ and using that 
\be
\tilde x_{\rm max}^0 = \left(\frac{r_0 \log N}{\sqrt{4 D t}}\right)^{1/2} = g^{1/2},
\ee 
one predicts
\bea \label{res} 
 &x_{\rm max} &\simeq \sqrt{4 D t \log N} + \sqrt{\frac{D t}{\log N}} \left( G  + h^{\rm KPZ}(0,g^2) + O(\log t) \right) \\
&& \simeq r_0 \sqrt{g} \left(\frac{4 D t}{r_0^2}\right)^{3/4} + \frac{r_0}{2 \sqrt{g}} \left(\frac{4 D t}{r_0^2}\right)^{1/4} 
\left( G  + h^{\rm KPZ}(0,g^2) + O(\log t) \right) .
\eea 
At fixed $g$ the leading behavior of the position of the maximum (the first term in \eqref{res}, which is deterministic) behaves as $x_{\rm max} \propto t^{3/4}$ in this moderate deviation regime. 
The fluctuations are $O(t^{1/4})$ and are given by the sum of two random variables,
i.e. the law of the sum is a convolution of a Gumbel distribution and of the PDF of
the height of the droplet solution of the KPZ equation at effective time $g^2$.
At fixed $g$ of order unity the Gumbel ``thermal'' fluctuations (different particles in the same environment)
are of the same order as the sample to sample fluctuations governed by the
KPZ equation. The latter exhibit a crossover from being Gaussian and $\propto g^{1/4}$ for $g \ll 1$,
to being large and $\propto g^{1/3}$ for $g \gg 1$ where they smoothly match the
the TW result in \eqref{tw}. It is important to stress that, in the limit of large $t$, the present regime, i.e. $\log N \propto \sqrt{t}$, is attained for more moderate values of $N$ 
than the large deviation regime $\log N \propto t$. Hence it should be easier to detect in experiments.\\ 

The exact solution for the model in Section \ref{sec:LeJanRaimondflow} gives in addition
the precise value of the deterministic $O(\log t)$ term mentioned above. From
\eqref{eq:convergenceexplicit} we obtain 
\be
{\mathsf P}(x_{\rm max} < 2^{1/4} \sqrt{g} t^{3/4} + y t^{1/2}) \simeq \exp\left(- e^{- 2^{1/4} \sqrt{g} t^{1/4} y 
- \frac{1}{4} \log \frac{t}{2 g^2} + h_t } \right),
\ee
where $h_t$ is a random variable converging in law to $h^{\rm KPZ}(0,g^2)$. This leads to
\be 
\label{eq:resmoreprecise}
x_{\rm max} = 2^{1/4} \sqrt{g} t^{3/4} + \frac{1}{2^{1/4} \sqrt{g}} t^{1/4} 
\left( G  + h^{\rm KPZ}(0,g^2) - c_N + \log \sqrt{2 \pi t}  \right) .
\ee
These results are in agreement with \eqref{res}, if one sets $r_0=2$, $D=1/2$,
and predict in addition the $O(\log t)$ term which was not explicit in \eqref{res} (recall that $c_N$ was defined in \eqref{gumb}).

\begin{remark}
The distribution of the random variable in \eqref{res} and \eqref{eq:resmoreprecise} also appears in an apparently unrelated context of fermions in a trap at finite temperature \cite{FermionsFiniteT}. 
\end{remark}

\appendix 

\renewcommand\thesection{Appendix \Alph{section}.}
\renewcommand\thesubsection{\Alph{section}.\arabic{subsection}}
\renewcommand\theremark{\Alph{section}.\arabic{remark}}

\section{More general continuous model}
\label{sec:interpolation}

The equation \eqref{eq:trueevolution} can be -- quite formally \cite[Section 6]{barraquand2019large} -- associated to the stochastic PDE 
\begin{equation}
\partial_t Z = \frac 1 2 \partial_x^2 Z + \frac{1}{\sqrt{\lambda}}\xi \partial_x Z, 
\label{eq:formalSPDE}
\end{equation} 
in the sense that the mixed moments $\mathbb E\left[ Z(x_1 ,t) \dots Z(x_k,t) \right]$ solve \eqref{eq:trueevolution} when one applies naively the definition of the covariance of the white noise $\mathbb E[\xi(x,t)\xi(s,y)] = \delta(x-y)\delta(t-s)$. 
The reason why this cannot be more than a formal computation is because one expects that a solution of \eqref{eq:formalSPDE} with initial data $Z_0(x)$ satisfies 
\begin{equation}
Z(x,t) = p_t \ast Z_0+ \int_{\R} dy \int_{0}^t ds \, p_{t-s}(x-y) \xi(y,s) \partial_y Z(y,s), 
\label{eq:Duhamelform}
\end{equation}
where $p_t(x)$ is the standard heat kernel and $\ast$ denotes the convolution of functions (in the space variable). The integral in \eqref{eq:Duhamelform} cannot be directly understood as an It\^o integral: indeed, there is a priori no reason for  $\partial_y Z(y,s)$ to be a function adapted to the filtration generated by $\xi(y',s') $ for $y'\in\R, s'<s$. The term $\partial_y Z(y,s)$ should be understood as a distribution and making sense of \eqref{eq:Duhamelform} requires to define the  product of two distributions $\partial_y Z(y,s)$ and $\xi(y,s)$.

\subsection{Interpolation between multiplicative SHE and stochastic flows} 
\label{sec:simplechangesofvariables}

Discarding the mathematical issues to make sense of \eqref{eq:formalSPDE} that we have just explained, let us  study the stochastic PDE  
\begin{equation}
\partial_t Z = \partial_x^2 Z + \xi (g_0 + g_1 \partial_x) Z,
\label{eq:spdemixed} 
\end{equation}
 where $g_0$ and $g_1$ are two parameters and $\xi(x,t)$ a unit space time white noise.
  It interpolates between \eqref{eq:formalSPDE} 
for $g_0=0$ and the multiplicative noise stochastic equation \eqref{she} for $g_1=0$
\footnote{ An interesting and more general interpolation reads
		$\partial_t Z = \partial_x^2 Z + g_0 \xi_0 + g_1 \xi_1 \partial_x Z$,
		where $\xi_0$ and $\xi_1$ are two Gaussian noises 
		with an arbitrary mutual correlation. }.
		Through the Cole-Hopf transformation, it is associated to a variant of the KPZ equation
with additional gradient noise
\begin{equation}
\partial_t h = \partial_x^2 h + (\partial_x h)^2  +  \xi g_1 \partial_x h  + g_0 \xi .
\end{equation}
Note that there is a particular solution of the equation \eqref{eq:spdemixed} which reads
\begin{equation}
Z_d(x,t) = e^{\frac{g_0^2}{g_1^2} t - \frac{g_0}{g_1} x} 
\end{equation}
and which does not see the noise since $(g_0 + g_1 \partial_x) Z_d=0$.
It is thus natural to write 
\begin{equation}
Z(x,t) = Z_d(x,t) \tilde Z(x,t),
\end{equation}
leading, for $g_1\neq 0$,  to the equation 
\begin{equation}
\partial_t \tilde Z = \partial_x^2 \tilde Z + \xi g_1 \partial_x \tilde Z  - \frac{2 g_0}{g_1} \partial_x \tilde Z .
\end{equation}
This can be solved as
\begin{equation}
\tilde Z(x,t) = \hat Z(x - 2 \frac{g_0}{g_1} t,t) 
\end{equation}
where $\hat Z$ satisfies \eqref{eq:formalSPDE} (without the prefactor $1/2$ in front of the Laplacian), that is 
\begin{equation}
\partial_t \hat Z = \partial_x^2 \hat Z + \hat \xi g_1 \partial_x \hat  Z  ,
\end{equation}
where $\hat \xi(x,t)=\xi(x+2 \frac{g_0}{g_1} t,t)$  is also a unit space time white noise 
if $\xi$ is one. Thus, we have found that as long as $g_1\neq 0$, the model associated with the stochastic PDE \eqref{eq:spdemixed} reduces, after changes of variables and multiplication by $Z_d$, to the stochastic PDE \eqref{eq:formalSPDE} associated with the integrable model of stochastic flows (considered in Section \ref{sec:LeJanRaimondflow}).

\subsection{Bose gas with interaction $g_0\delta(x_i-x_j) + g_1\delta(x_i-x_j)\partial_{x_i} \partial_{x_j}$}

 The mixed moments $\mathbb E [Z(x_1,t ) \dots Z(x_k,t)]$ of the solution $Z(x,t)$ of the equation \eqref{eq:spdemixed} solve the equation 
\begin{equation}
\partial_t u = \sum_i \partial_{x_i}^2 u + \sum_{i \neq j} \delta(x_i-x_j) (g_1 \partial_{x_i} + g_0) 
(g_1 \partial_{x_j} + g_0) u.
\label{eq:operatormixedinteraction}
\end{equation}
According to Section \ref{sec:simplechangesofvariables}, the above equation \eqref{eq:operatormixedinteraction} should reduce to \eqref{eq:trueevolution} after appropriate changes of variables and renormalization. In this section, we show that we may as well solve \eqref{eq:operatormixedinteraction} directly by Bethe ansatz.  A similar operator was considered in \cite{stouten2018something} in the context of quantum integrable model for interacting bosons,
and it was noticed that such operators should be Bethe ansatz diagonalizable (in \cite{stouten2018something}, the partial derivatives and delta functions are exchanged, this would correspond to look at forward transition probabilities for our diffusion models, instead of backward transition probabilities). 

\bigskip 
The operator applied to $u$ on the r.h.s of \eqref{eq:operatormixedinteraction} preserves the symmetry with respect to permuting variables. As we are interested in this paper in solutions that are symmetric (the mixed moments are definitely symmetric with respect to permuting variables $x_i$'s)  we can restrict our considerations to the set of symmetric functions. Hence we only need to determine the solution $u$ on the sector $x_1\geqslant \dots \geqslant x_k$. The boundary condition when $x_i=x_{i+1}$ is
\begin{equation}
(\partial_{x_i} - \partial_{x_{i+1}} + g_1^2 \partial_{x_i}  \partial_{x_{i+1}} + g_0^2 
+ g_0 g_1 (\partial_{x_i}  + \partial_{x_{i+1}}  ) ) u = 0.
\label{eq:boundaryconditionmixed}
\end{equation}
For certain initial data,  we may search for solutions of the form
\begin{equation}
u(\vec x, t) = \int_{r_1+\I\R} \frac{dz_1}{2\I\pi} \dots \int_{r_k+\I\R}\frac{dz_k}{2\I\pi} \prod_{i=1}^k g(z_i) e^{ f(z_i) x_i + f(z_i)^2 t } \prod_{A<B} \frac{z_A-z_B}{z_A-z_B-1}, 
\label{eq:solutionwithvariablez}
\end{equation}
on the sector $x_1\geqslant \dots \geqslant x_k$, where the variable  $z_i$ is integrated along the vertical line $r_i+\I\R$ where for $i>j$, $r_i>r_j+1$, and the function $g$ is related to the initial data for $u$. 

In order for the boundary condition \eqref{eq:boundaryconditionmixed} to be satisfied, the function $f(z)$ must be such that 
\begin{equation}
f(z_1) - f(z_2) + g_1^2 f(z_1) f(z_2) + g_0^2 + g_0 g_1 (f(z_1)+f(z_2))
\label{eq:equationforf}
\end{equation}
evaluated at $z_1=z_2+1$ vanishes. In that case, the left hand side of \eqref{eq:boundaryconditionmixed} can be written as a contour integral where there is no more pole at $z_i=z_{i+1}+1$, so that the contours for $z_{i+1}$ can be deformed to be the same as the contour for $z_i$ (that is $r_i+\I\R$), and the integral vanishes because the integrand is antisymmetric with respect to exchanging $z_i,z_{i+1}$. 

We found two family of solutions of \eqref{eq:equationforf}, which we show are in fact the same:
\begin{equation}
f(z) = \frac{-g_0^2 (z + \gamma)}{1+ g_0 g_1(z+ \gamma)},
\label{eq:choiceforf1}
\end{equation}
and
\begin{equation}
f(z) =  \frac{1 - g_0 g_1 (z + \delta)}{g_1^2 (z+\delta)}, 
\label{eq:choiceforf2}
\end{equation}
which are identical upon the relation
\begin{equation}
\delta- \gamma   = \frac{1}{g_0 g_1}.
\end{equation}

In general, eigenfunctions of \eqref{eq:operatormixedinteraction} can be written (in the sector $x_1\geqslant \dots \geqslant x_k$) as 
\begin{equation}
\Psi_z (\vec x)= \sum_{\sigma\in S_n} \sigma\left(  \prod_{A<B} \frac{z_A-z_B-1}{z_A-z_B} \prod_i e^{ f(z_i) x_i } \right),
\end{equation}
where  the notation $\sigma(\cdot)$ means that for any function $F$, $\sigma\left(F(\vec z)\right) = F(z_{\sigma(1)}, \dots, z_{\sigma(k)})$. We expect that these functions $\Psi_{\vec z}$ are limits of so-called spin-Hall-Littlewood functions, introduced in \cite{borodin2017family} (see also \cite{borodin2016higher}), which are rational deformations of the Hall-Littlewood symmetric functions. Likewise, the functions  $\Psi_{\vec z}$ are rational deformations of the Bethe ansatz eigenfunctions of the Lieb-Liniger model (the rational transformation is given by \eqref{eq:choiceforf1}, \eqref{eq:choiceforf2} and is similar with rational transformations considered in \cite{stouten2018something}).

In order to discuss the initial data, it is convenient to use the change of variable $f(z)=w$, or equivalently we can look directly for a solution 
of the following form form: for
$x_1\geqslant x_2\geqslant  \dots \geqslant  x_k$, 
\begin{equation}
u(x_1,\dots x_n, t)=\int_{\alpha_1+\I\R}\frac{dw_1}{2\I\pi} \dots \int_{\alpha_k+\I\R} \frac{dw_k}{2\I\pi} \prod_{i=1}^k \frac{g_0}{g_0+g_1 w_i}  e^{ w_i x_i + w_i^2 t } \prod_{a<b} 
\frac{w_b-w_a}{w_b-w_a-  (g_0 + g_1 w_a)(g_0+g_1 w_b)} 
\label{eq:solutionwithvariablew}
\end{equation}
where the contours are vertical lines such that $-g_0/g_1<\alpha_1\ll \dots \ll \alpha_k$ (the $\alpha_i$ are sufficiently spaced so that the poles for the variable $w_b$ in the product over $a<b$ all lie on the left of $\alpha_b+\I\R$). 
The formula \eqref{eq:solutionwithvariablew} solves \eqref{eq:operatormixedinteraction} \eqref{eq:boundaryconditionmixed} for the initial condition 
\begin{equation}
u(\vec x , 0) = \prod_{i=1}^k\frac{g_0}{g_1} e^{-g_0 x_i/g_1} \theta(x_i), \quad \text{ where } x_i\neq 0 \text{ for all }1\leqslant i\leqslant k,
\label{eq:initialcondition}
\end{equation}
and $\theta(x)=\mathds{1}_{x>0}$.
Indeed, at $t=0$, if $x_k<0$ then we may shift horizontally the contour for $w_k$ to $+\infty$, and the integrand uniformly converges to zero because of the factor $e^{w_kx_k}$. Hence we may assume that $x_1\geqslant \dots\geqslant x_k\geqslant 0$.  Now we may shift the contour for  $w_1$ to $-\infty$. if $x_1>0$, the integral will be zero for the same reason as previously, but we have crossed a pole during the contour deformation at $w_1=-g_0/g_1$. The associated residue is readily computed and it equals $g_0/g_1 e^{g_0x_1/g_1}$. One may then proceed to the same contour deformation with all variables successively and obtain the claimed formula. If some $1\leqslant i \leqslant k$ is such that $x_k=x_{k-1} = \dots = x_i=0$, the integral can be computed but \eqref{eq:initialcondition} does not hold. 

Going back to \eqref{eq:solutionwithvariablez}, the function $g(z)$ is determined from \eqref{eq:solutionwithvariablew} using the Jacobian
\begin{equation}
dz g(z) = dw \frac{g_0}{g_0+ g_1 w}. 
\end{equation}

\begin{remark}
Now that we have found $f(z)$ and $g(z)$ we may apply e.g.  \cite[Proposition 6.2.7]{borodin2014macdonald} \eqref{eq:solutionwithvariablez} so as to write the solution as a sum of integrals of determinants. These integrals of determinants may be further assembled to form Fredholm determinants convenient for asymptotic analysis. We will not discuss this further, since we have observed in Section \ref{sec:simplechangesofvariables} that the model corresponding to \eqref{eq:operatormixedinteraction} or \eqref{eq:spdemixed} can be reduced to the model studied in \cite{barraquand2019large} for which determinantal formulas have been already proved. 
\end{remark}


\begin{thebibliography}{12}
	    \bibitem{barraquand2017random}
		G.~Barraquand and I.~Corwin.
		\newblock Random-walk in beta-distributed random environment.
		\newblock {\em Probab. Theory Rel. Fields}, 167(3):1057--1116, Apr 2017. arXiv:1503.04117.
		
		\bibitem{PLDTTDiffusion}
		P. Le Doussal, T. Thiery,
		{\it Diffusion in time-dependent random media and the Kardar-Parisi-Zhang equation}, 
		arXiv:1705.05159, Phys. Rev. E 96, 010102 (2017).
		
		\bibitem{KPZ}
		M. Kardar, G. Parisi and Y.C. Zhang,
		{\it Dynamic Scaling of Growing Interfaces},
		Phys. Rev. Lett. {\bf 56}, 889 (1986).
		
		\bibitem{corwin2017kardar}
		I.~Corwin and Y.~Gu.
		\newblock {Kardar--Parisi--Zhang equation and large deviations for random walks
			in weak random environments}.
		\newblock {\em J. Stat. Phys.}, 166(1):150--168, 2017. arXiv:1606.07332.
		
		\bibitem{barraquand2019large}
		G.~Barraquand and M.~Rychnovsky.
		\newblock Large deviations for sticky Brownian motions.
		\newblock {\em arXiv preprint arXiv:1905.10280}, 2019.
		
		\bibitem{rassoul2009almost}
		F.~Rassoul-Agha and T.~Sepp{\"a}l{\"a}inen.
		\newblock Almost sure functional central limit theorem for ballistic random
		walk in random environment.
		\newblock {\em Ann. Inst. Henri Poincar{\'e} Probab. Stat}, 45(2):373--420,
		2009.
		
		\bibitem{RAP} 
		M. Bal\'azs, F. Rassoul-Agha, T. Sepp\"al\"ainen,
		{\it The random average process and random walk in a space-time random environment in one dimension},
		Commun. Math. Phys. {\bf 266} 499, 2006.
		
		\bibitem{Yu16}
		J.~Yu.
		\newblock {Edwards--Wilkinson fluctuations in the Howitt--Warren flows}.
		\newblock {\em Stoch. Proc. Appl.}, 126(3):948--982, 2016.
		
		\bibitem{LargeDev}
		F. Rassoul-Agha, T. Sepp\"al\"ainen, A. Yilmaz,
		{\it Quenched free energy and large deviations for random walks in random potentials},
		Commun. Pure Appl. Math., {\bf 66} 202244 (2013). arXiv:1104.3110.
		
		\bibitem{povolotsky2013integrability}
		A.~M. Povolotsky.
		\newblock On the integrability of zero-range chipping models with factorized
		steady states.
		\newblock {\em J. Phys. A}, 46(46):465205, 2013.
		
		\bibitem{usBeta}
		T. Thiery, P. Le Doussal,
		{\it Exact solution for a random walk in a time-dependent 1D random environment: the
			point-to-point Beta polymer},
		Journal of Physics A: Mathematical and Theoretical {\bf 50} 4 (2016). arXiv:1605.07538.
		
		\bibitem{balazs2019large}
		M.~Bal{\'a}zs, F.~Rassoul-Agha, and T.~Sepp{\"a}l{\"a}inen.
		\newblock Large deviations and wandering exponent for random walk in a dynamic
		beta environment.
		\newblock {\em Ann. Probab.}, 47(4):2186--2229, 2019.
		
		\bibitem{le2004flows}
		Y.~Le~Jan and O.~Raimond.
		\newblock {Flows, coalescence and noise}.
		\newblock {\em Ann. Probab.}, 32(2):1247--1315, 2004.
		
		\bibitem{le2004sticky}
		Y.~Le~Jan and O.~Raimond.
		\newblock Sticky flows on the circle and their noises.
		\newblock {\em Probab. Theory Related Fields}, 129(1):63--82, 2004.
		
		\bibitem{le2004products}
		Y.~Le~Jan and S.~Lemaire.
		\newblock Products of {B}eta matrices and sticky flows.
		\newblock {\em Probab. Theory Rel. Fields}, 130(1):109--134, 2004.
		

		
		\bibitem{Howitt2009consistent}
		C.~Howitt and J.~Warren.
		\newblock {Consistent families of Brownian motions and stochastic flows of
			kernels}.
		\newblock {\em Ann. Probab.}, 37(4):1237--1272, 2009.

		\bibitem{schertzer2014stochastic}
		E.~Schertzer, R.~Sun, and J.~M. Swart.
		\newblock {Stochastic flows in the Brownian web and net}.
		\newblock {\em Memoirs Amer. Math. Soc.}, 227(1065):1--172, 2014.
		
		\bibitem{schertzer2015brownian}
		E.~Schertzer, R.~Sun, and J.~M. Swart.
		\newblock {The Brownian web, the Brownian net, and their universality}.
		\newblock {\em Advances in Disordered Systems, Random Processes and Some
			Applications}, pages 270--368, 2015. arXiv:1506.00724.
		
			\bibitem{stouten2018something}
		E.~Stouten, P.~W. Claeys, M.~Zvonarev, J.~Caux, and V.~Gritsev.
		\newblock Something interacting and solvable in 1d.
		\newblock {\em J. Phys. A: Math. Theor.}, 51(48):485204, 2018.
		
		\bibitem{kardareplica} M. Kardar, Nucl. \textit{Replica Bethe ansatz studies of two-dimensional interfaces with quenched random impurities}, Phys. B {\bf 290}, 582 (1987).
		
		\bibitem{gawkedzki2004sticky}
		K.~Gawedzki and P.~Horvai.
		\newblock {Sticky behavior of fluid particles in the compressible Kraichnan
			model}.
		\newblock {\em J. Stat. Phys.}, 116(5-6):1247--1300, 2004.
		
		\bibitem{warren2015sticky}
		J.~Warren.
		\newblock Sticky particles and stochastic flows.
		\newblock In {\em In Memoriam Marc Yor-S{\'e}minaire de Probabilit{\'e}s
			XLVII}, pages 17--35. Springer, 2015.

		\bibitem{kraichnan1968small}
		R.~H. Kraichnan.
		\newblock {Small-scale structure of a scalar field convected by turbulence}.
		\newblock {\em The Physics of Fluids}, 11(5):945--953, 1968.
		
		
				\bibitem{gawedzki1995anomalous}
				K.~Gawedzki and A.~Kupiainen.
				\newblock {Anomalous scaling of the passive scalar}.
				\newblock {\em Phys. Rev. Lett.}, 75(21):3834, 1995.
		
				\bibitem{gawedzki1996university}
				K.~Gawedzki and A.~Kupiainen.
				\newblock {University in turbulence: An exactly solvable model}.
				\newblock In {\em Low-dimensional models in statistical physics and quantum
					field theory}, pages 71--105. Springer, 1996.
				
				\bibitem{bernard1998slow}
				 D. Bernard, K. Gawedzki, A. Kupiainen, 
					\newblock Slow modes in passive advection.
					\newblock {\em J. Stat. Phys.}, 90.3-4 (1998): 519-569, arXiv:cond-mat/9706035.
		
				\bibitem{gawedzki2000phase}
				K.~Gawedzki and M.~Vergassola.
				\newblock {Phase transition in the passive scalar advection}.
				\newblock {\em Phys. D: Nonlinear Phenomena}, 138(1-2):63--90, 2000.

		
		
		\bibitem{feller1952parabolic}
		W.~Feller.
		\newblock {The parabolic differential equations and the associated semi-groups
			of transformations}.
		\newblock {\em Ann. Math.}, pages 468--519, 1952.
		
		
		\bibitem{arratia1979coalescing}
		R.  Arratia.  \textit{Coalescing  Brownian  motions  on  the  line}.  Ph.D.  Thesis,  University  of Wisconsin, Madison, 1979.
		
		\bibitem{toth1998true}
		B.  T\'oth  and  W.  Werner. \textit{ The  true  self-repelling  motion}, Probab.  Theory  Related Fields, 111:375–452, 1998.
		
		\bibitem{sun2008brownian}
		R.~Sun and J.~M. Swart.
		\newblock {The {B}rownian net}.
		\newblock {\em Ann. Probab.}, 36(3):1153--1208, 2008.
		
		\bibitem{newman2010marking}
		C.M. Newman, K. Ravishankar, and E. Schertzer. \textit{Marking (1,2) points of the Brownian web and applications.} Ann.  Inst.  Henri  Poincar\'e  Probab.  Statist. 46, 537–574, 2010.
		
		\bibitem{fontes2004brownian} 
		L.R.G. Fontes, M. Isopi, C.M. Newman, K. Ravishankar. The Brownian web: characterization and convergence. Ann. Probab.32(4), 2857–2883, 2004.
		


		
		\bibitem{schertzer2009special}
		E.~Schertzer, R.~Sun, and J.~M. Swart.
		\newblock {Special points of the {B}rownian net}.
		\newblock {\em Electr. J. Probab.}, 14:805--864, 2009.
		
		\bibitem{le2002integration}
		Y.~Le~Jan and O.~Raimond.
		\newblock {Integration of Brownian vector fields}.
		\newblock {\em Ann. Probab.}, 30(2):826--873, 2002.

		
		
		\bibitem{borodin2014macdonald}
		A.~Borodin and I.~Corwin.
		\newblock Macdonald processes.
		\newblock {\em Probab. Theory and Rel. Fields}, 158(1-2):225--400, 2014.
		
		\bibitem{ghosal2018moments}
		P.~Ghosal.
		\newblock Moments of the {SHE} under delta initial measure.
		\newblock {\em arXiv preprint arXiv:1808.04353}, 2018.
		

		
		
		\bibitem{FermionsFiniteT} 
		D. S. Dean, P. Le Doussal, S. N. Majumdar, G. Schehr, Non-interacting fermions at finite temperature in a d-dimensional trap: universal correlations,
		arXiv:1609.04366, Phys. Rev. A 94, 063622 (2016).
		
	
		
		
		\bibitem{borodin2017family}
		A.~Borodin.
		\newblock On a family of symmetric rational functions.
		\newblock {\em Adv. Math.}, 306:973--1018, 2017.
		
		\bibitem{borodin2016higher}
		A.~Borodin and L.~Petrov.
		\newblock Higher spin six vertex model and symmetric rational functions.
		\newblock {\em Selecta Math.}, 24(2):751--874, 2018.
\end{thebibliography}
\end{document}